\documentclass[aps,rmp,eqsecnum,amssymb,12pt]{article}
\usepackage{graphicx}
\usepackage{amssymb}
\usepackage{amsmath}
\usepackage[dvips]{color}

\def\IR{\relax{\rm I\kern-.18em R}}
\def\beq{\begin{equation}}
\def\eeq{\end{equation}}

\def\beqa{\begin{eqnarray}}
\def\eeqa{\end{eqnarray}}

\begin{document}

\begin{titlepage}

\setcounter{page}{1} \baselineskip=15.5pt \thispagestyle{empty}

\begin{flushright}
PUPT-2241\\
\end{flushright}
\vfil

\begin{center}
{\LARGE Entanglement as a Probe of Confinement}
\end{center}
\bigskip\

\begin{center}
{\large Igor R. Klebanov,$^{a}$ David Kutasov,$^{b}$ Arvind Murugan$^{a}$}
\end{center}

\begin{center}
\textit{${}^a$ Department of Physics and Center for Theoretical Physics,
Princeton University \\ Princeton, NJ 08544 U.S.A \\
${}^b$ Department of Physics, University of Chicago \\ 5640 S.Ellis Av., Chicago, IL 60637 U.S.A }
\end{center} \vfil


\noindent  We investigate the entanglement entropy in gravity duals
of confining large $N_c$ gauge theories using the proposal of \cite{Tak}. Dividing one of the
directions of space into a line segment of length $l$ and its
complement, the entanglement entropy between the two subspaces is
given by the classical action of the minimal bulk
hypersurface which approaches the endpoints of the line segment at
the boundary. We find that in confining backgrounds there are
generally two such surfaces. One consists of two disconnected
components localized at the endpoints of the line segment. The other
contains a tube connecting the two components. The disconnected
surface dominates the entropy for $l$ above a certain critical value
$l_{\rm crit}$ while the connected one dominates below that value.
The change of behavior at $l=l_{\rm crit}$ is reminiscent of the
finite temperature deconfinement transition: for $l < l_{\rm crit}$
the entropy scales as $N_c^2$, while for $l > l_{\rm crit}$ as
$N_c^0$. We argue that a similar transition should occur in any
field theory with a Hagedorn spectrum of non-interacting bound
states. The requirement that the entanglement entropy has
a phase transition may be useful in constraining gravity duals of confining theories.

\vfil

\end{titlepage}

\newpage

\tableofcontents

\pagestyle{headings}

\section{Introduction}

Consider a $d+1$ dimensional quantum field theory (QFT) on
$\IR^{d+1}$ in its vacuum state $|0\rangle$. Divide the $d$
dimensional space into two complementary regions,
\beqa\label{twopp}
&A&=\IR^{d-1}\times I_l\ ,\nonumber \\
&B&=\IR^{d-1}\times(\IR- I_l) \ ,
\eeqa
where $I_l$ is a line segment of length $l$. The entanglement
entropy between the regions $A$ and $B$ is defined as the entropy
seen by an observer in $A$ who does not have access to the degrees
of freedom in $B$, or vice versa (see e.g. \cite{Calabrese:2005zw}
for a recent review and references to earlier work). It can be
calculated by tracing the density matrix of the vacuum,
$\rho_0=|0\rangle\langle 0|$, over the degrees of freedom in $B$ and
forming the reduced density matrix
\beq
\label{redmat} \rho_A={\rm Tr}_B \rho_0~.
\eeq
The quantum entanglement entropy $S_A$ is then given by
\beq
\label{entdef} S_A=-{\rm Tr}_A \rho_A\ln \rho_A~.
\eeq
The above construction can be generalized in a number of ways. In
particular, one can replace the vacuum state $|0\rangle$ by any
other pure or mixed state,  and choose the submanifold of $\IR^d$,
$A$,  to be different than (\ref{twopp}). In this paper we will
restrict to the choices above, which are sufficient for our
purposes.

The entanglement entropy (\ref{entdef}) is in general  UV divergent.
To leading order in the UV cutoff $a$ it scales like
\cite{Bombelli:1986rw,Srednicki:1993im}
\beq \label{leaddiv} S_A\simeq {V_{d-1}\over a^{d-1}}
\eeq
where $V_{d-1}$ is the volume of $\IR^{d-1}$ in (\ref{twopp}). Note
that (\ref{leaddiv}) is independent of $l$. This turns out to be a
general feature -- the entropy is defined up to an $l$ independent
(infinite) additive constant. In particular, $\partial_l S_A$ and
differences of entropies at different values of $l$ approach a
finite limit as $a\to 0$. In $d+1$ dimensional CFT with $d>1$, the
finite $l$-dependent part of the entropy is negative and
proportional to ${V_{d-1}/l^{d-1}}$, while for $d=1$ it goes like
$\ln l$.

If the QFT in question has a gravity dual \cite{AdS-CFT}, it is
natural to ask whether the entanglement entropy can be calculated
using the bulk description. This problem was addressed in
\cite{Tak}. For the case of $d+1$ dimensional large $N_c$ conformal
field theories with $AdS_{d+2}$ gravity duals, the authors of
\cite{Tak} proposed a simple geometric method for computing the
entanglement entropy and subjected it to various tests. This method
is to find the minimal area $d$-dimensional surface $\gamma$ in
$AdS_{d+1}$ such that the boundary of $\gamma$ coincides with the
boundary of $A$, which in the case (\ref{twopp}) consists of two
copies of $\IR^{d-1}$ a distance $l$ apart. The quantum entanglement
entropy between the regions $A$ and $B$ is proportional to the
classical area of this surface,
\beqa\label{entropy} S_A = \frac{1}{4G_N^{(d+2)}} \int_{\gamma} d^d \sigma
\sqrt{G^{(d)}_{\rm ind}}~,
\eeqa
where $G_N^{(d+2)}$ is the $d+2$ dimensional Newton constant and
$G^{(d)}_{\rm ind}$ is the induced string frame metric on $\gamma$.
Note that the surface $\gamma$ is defined at a fixed time and
(\ref{entropy}) gives the entanglement entropy at that time. For
static states, such as the vacuum, the resulting entropy is time
independent.\footnote{Generalizations of the proposal of \cite{Tak}
to time dependent states were discussed in \cite{Hubeny:2007xt}.}
Also, since $\gamma$ is extended in the transverse $\IR^{d-1}$
(\ref{twopp}), the entropy (\ref{entropy}) is proportional to its
volume $V_{d-1}$. Thus, in this case it is better to consider the
entropy per unit transverse volume.

In non-conformal  theories, the volume of the $8-d$ compact
dimensions and the dilaton are in general not constant. A natural
generalization of (\ref{entropy}) to the corresponding ten
dimensional geometries is \cite{Tak,Nishioka}
\beqa \label{prescription} S_A = \frac{1}{4G_N^{(10)} } \int d^8 \sigma e^{-2
\phi} \sqrt{G^{(8)}_{\rm ind}}~.
\eeqa
The entropy is obtained by minimizing the action
(\ref{prescription}) over all surfaces that approach the boundary of
$A$ (\ref{twopp}) at the boundary of the bulk manifold, and are
extended in the remaining spatial directions. Since $ G_N^{(10)}=
8\pi^6 \alpha'^4 g_s^2$, this gives an answer of order $N_c^2$ in
the 't Hooft limit $N_c\to\infty$ with $g_s N_c$ held fixed.

It was shown in \cite{Tak} that for $AdS_3$ the prescription
(\ref{entropy}) successfully reproduces the known form of the
entanglement entropy in two-dimensional conformal field theory, and
that it gives sensible results when applied to some higher
dimensional vacua, such as $AdS_5\times S^5$. Nevertheless, some
aspects of the proposal are not well understood. In particular, it
is not clear how to extend it beyond leading order in $1/N_c$.

In this paper we apply the proposal of \cite{Tak,Nishioka} to confining
backgrounds, such as \cite{Witten,KS}. One of the motivations for this investigation is to
subject the proposal (\ref{prescription}) to further tests. Another
is to study the $l$ dependence of the entanglement entropy, which is
in general difficult to determine in strongly coupled field
theories.

Gravitational backgrounds dual to confining gauge theories typically
have the following structure. As one moves in the radial direction
away from the boundary, an internal cycle smoothly contracts and approaches
zero size at the infrared (IR) end of space. The radial direction
together with the shrinking cycle make a type of cigar geometry,
with the IR end of space corresponding to the tip of the cigar.

We will see that in such geometries there are in general multiple
local minima of the action (\ref{prescription}) for given $l$. One
of those is a disconnected surface, which consists of two cigars
extended in $\IR^{d-1}$ and separated in the remaining direction in
$\IR^d$ by the distance $l$. A second one is a connected surface, in
which the two cigars are connected by a tube whose width depends on
$l$. Since the two geometries are related by a continuous
deformation, there is a third extremum of the action between them,
which is a saddle point of (\ref{prescription}). A natural
generalization of the proposal of \cite{Tak} to this case is to
identify the entanglement entropy with the absolute minimum of the
action. We will see that this leads to a phase transition in the
behavior of the entanglement entropy as a function of $l$. 
\footnote{
This phenomenon has already been noted in one specific 
example \cite{Nishioka} -- the static ``AdS bubble,'' 
which is equivalent to the background of D3-branes on a 
circle that we study in section 4 (we thank T. Takayanagi 
for pointing this out to us).}

For the disconnected solution, $S_A$ (\ref{prescription}) does not
depend on $l$. As mentioned above, the actual value of $S_A$ depends
on the UV cutoff, but if we are only interested in differences of
entropies, or the derivative of the entropy with respect to $l$, we
can set it to zero. For the connected solution, $S_A$ depends
non-trivially on $l$. For small $l$, it is smaller than that of the
disconnected one. Thus, it dominates the entropy
(\ref{prescription}). For $l>l_{\rm crit}$ the action of the
connected solution becomes larger than that of the disconnected one,
and it is the latter that governs the entropy. Thus, in going from
$l<l_{\rm crit}$ to $l>l_{\rm crit}$, $\partial_l S_A$ goes from
being of order $N_c^2$ to being of order $N_c^0$. One can think of
this change of behavior as a phase transition which, as we show, is
typical in large $N_c$ confining theories.

Similar transitions between connected and disconnected D-brane
configurations play a role in other contexts. In
\cite{Brandhuber} an analogous transition is responsible for
screening of magnetic charges in confining gravitational
backgrounds; in \cite{Giveon:2007fk} it governs the pattern of
metastable supersymmetry breaking vacua in a brane construction of
supersymmetric QCD. An important difference is that in all these
cases the transitions involve the rearrangement of real branes,
whereas the hypersurface whose area is being minimized here does not
seem to have such an interpretation.

The plan of the rest of the paper is as follows. In section 2 we
present a general analysis of a class of gravitational backgrounds
that arises in the construction of holographic duals of confining
gauge theories. We
show that in this class there are multiple local
minima of the action (\ref{prescription}), as discussed above.
With some mild assumptions, we
also show that for small $l$ the global
minimum of the action corresponds to a connected solution, while for
large $l$ it corresponds to a disconnected one. We also
show that the connected solution does not exist for sufficiently
large $l$.

In sections 3 -- 5 we illustrate the discussion of section 2 with a
few examples. Section 3 contains an analysis of the geometry of
$N_c$ $D4$-branes wrapped around a circle with twisted boundary
conditions for the fermions. For $g_s N_c \ll 1$ this system reduces
at low energies to pure Yang-Mills (YM) theory, while for $g_s N_c
\gg 1$ it can be analyzed using the near-horizon geometry of the
$D4$-branes \cite{Witten}. In section 4 we describe the analogous
$D3$-brane system, which for $g_s N_c \ll 1$ gives rise to YM in
$2+1$ dimensions. Some of the results of this section were
obtained already in \cite{Nishioka}. Section 5 contains an analysis of the warped
deformed conifold (KS) background \cite{KS}, which corresponds to a
cascading, confining $SU(M(k+1))\times SU(Mk)$ supersymmetric gauge
theory. This theory approaches pure $SU(M)$ SYM theory in the limit
$g_s M\ll 1$, while the dual supergravity description is reliable in
the opposite limit, $g_s M\gg 1$.

In section 6 we connect the results of sections 2 -- 5 to large
$N_c$ confining field theories such as pure YM. To leading order in
$1/N_c$, such theories are expected to reduce to free field theories
of the gauge singlet bound states. The latter are expected to have a
Hagedorn density of states at high mass, $\rho(m)\sim
m^\alpha\exp(\beta_H m)$. The entanglement entropy in such theories
can be calculated by summing the contributions of the individual
states. We show that this sum over states has a very similar
character to the finite temperature partition sum, with $l$ playing
the role of the inverse temperature $\beta$. It converges for
sufficiently large $l$ and diverges below a critical value of $l$,
since the large entropy overwhelms the exponential suppression of
the contribution of a given state of large mass. In the
thermodynamic case, this phenomenon is related to the appearance of
a deconfinement transition. By analogy, it is natural to expect that
here it signifies a transition between an entropy that goes like
$N_c^0$ at large $l$, and one that goes like $N_c^2$ below a
critical value. Since the gravitational analysis reproduces this
feature of the dynamics, we conclude that the
system with $g_s N_c\gg 1$
is in the same universality class as the one with $g_s N_c\ll 1$.

In section 7 we comment on our results and discuss other systems
which one can analyze using similar techniques. We also point out some
general issues related to the proposal of \cite{Tak}.

\section{Holographic computation of entropy}
\label{Generalsection} The gravitational backgrounds we will
consider have the string frame metric
\beqa\label{gengeom}
ds^2 =  \alpha(U) \left[ \beta(U) dU^2 + dx^\mu
dx_\mu \right] + g_{ij} dy^i dy^j
\eeqa
%
where $x^\mu$ ($\mu = 0,1,\ldots,d$) parametrize $\IR^{d+1}$, $U$ is
the holographic radial coordinate, and $y^i$ $(i=d+2,\cdots,9)$ are
the $8-d$ internal directions. The volume of the internal manifold,
\beqa V_{\rm int}  = \int \prod_{i=1}^{8-d} dy^i \sqrt{{\rm det}
g}~,
\eeqa
and the dilaton, $\phi$, are taken to depend only on $U$.

The radial coordinate $U$ ranges from a minimal value, $U_0$, to
infinity. As $U\to U_0$, a $p$-cycle in the internal
$(8-d)$-dimensional space shrinks to zero size, so $V_{\rm
int}(U_0)=0$. The vicinity of $U=U_0$ looks locally like the origin
of spherical coordinates in $\IR^{p+1}$ (times a compact space), and
we assume that all the supergravity fields are regular there. In
particular, $\alpha(U)$ and $\phi(U)$ approach fixed finite values
as $U\to U_0$. The fact that $\alpha(U_0)> 0$ implies that the
string tension is non-vanishing. This is the gravitational
manifestation of the fact that the dual gauge theory is confining.

Examples of backgrounds in this class that will be discussed below
are the geometries of coincident $D3$ and $D4$-branes on a circle
with twisted boundary conditions \cite{Witten}, in which the
shrinking cycle is a circle ($p=1$), and the KS geometry \cite{KS}
in which it is a two-sphere ($p=2$). In the $D3$-brane and KS cases,
the dilaton is independent of $U$.

We would like to use the proposal (\ref{prescription}) to calculate
the entanglement entropy of $A$ and $B$ (\ref{twopp}) in the
geometry (\ref{gengeom}). Denoting the direction along which the
line segment $I_l$ in (\ref{twopp}) is oriented by $x$, the entropy
per unit volume in the transverse $\IR^{d-1}$ is given by
\beqa\label{forment}
{S_A\over V_{d-1}} =
\frac{1}{4G_N^{(10)} }
\int_{-{l\over2}}^{l\over2}dx \sqrt{H(U)} \sqrt{1 + \beta(U)
(\partial_x U)^2}
\eeqa
where we introduced the notation
\beqa\label{denote} H(U) = e^{-4 \phi} V_{\rm int}^2  \alpha^d~.
\eeqa
Due to the shrinking of the $p$-cycle, we have $H(U_0)=0$. Thus, as
$U$ varies between $U_0$ and $\infty$, $H(U)$ varies between $0$ and
$\infty$. It provides a natural parametrization of the radial
direction of the space (\ref{gengeom}). Near $U_0$, one has $H\sim
r^{2p}$, where $r\in [0,\infty)$ is a natural radial coordinate,
$dr=\sqrt{\beta(U)} dU$.

The quantity (\ref{denote}) is simply related to the warp factor we
get upon dimensionally reducing on the $(8-d)$-dimensional compact
manifold. The resulting $(d+2)$-dimensional Einstein frame metric
may be written as
\beqa\label{redgeom} ds_{d+2}^2 =  \kappa (U) \left[ \beta(U) dU^2
+dx^\mu dx_\mu \right]~.
\eeqa
A standard calculation shows that $\kappa(U)^d = H (U)$. It is a
common assumption that the warp factor $\kappa(U)$ is a monotonic
function of the holographic radial coordinate. In particular,
finiteness of the holographic central charge \cite{Girardello},
\beqa c \sim  \beta^{d\over2}\kappa^{3 d \over2} \left (\kappa'\right
)^{-d} \ ,
\eeqa
requires $\kappa$ to be monotonic. Since it goes to zero as $U\to
U_0$ and to infinity as $U\to\infty$, it must be that $\kappa'>0$
for all $U$. This implies $H'(U)>0$, a fact that will be useful
below.

We need to find the shape $U(x)$ that minimizes the action
(\ref{forment}) subject to the constraint $U(x\to\pm
{l\over2})\to\infty$. Denoting by $U^*$ the minimal value of $U$
along this curve,\footnote{If the curve is smooth, this value is
attained at $x=0$, where $\partial_xU=0$.} and using the fact that
the action does not depend directly on $x$, its equation of motion
can be integrated once and written in the form
\beqa\label{firstord}
\partial_x U =\pm \frac{1}{\sqrt{\beta}} \sqrt{\frac{H(U)}{H(U^*)} -
1}~.
\eeqa
Integrating once more we find
\beqa \label{generall} l(U^*) = 2\sqrt{H(U^*)}\int_{U^*}^\infty
\frac{dU\sqrt{\beta(U)}} {\sqrt{ H(U) -  H(U^*)}}~. \eeqa
Plugging (\ref{firstord}) into (\ref{forment}) we find
\beqa \label{generalS} \frac{S_A}{V_{d-1}} = \frac{1}{2G_N^{(10)} }
\int_{U^*}^{U_\infty} \frac{dU\sqrt{\beta(U)}H(U)
}{\sqrt{H(U)-H(U^*)}}~.
\eeqa
In the examples we study below, and probably much more
generally, the integral in (\ref{generall}) turns out to be
convergent, while that in (\ref{generalS}) is not. This is the
reason for the appearance of the UV cutoff $U_\infty$ in the latter
and its absence in the former.

As mentioned earlier, the entropy $S_A$ depends on the cutoff only
via an $l$ independent constant, which cancels in differences of
entropies. This can be seen from (\ref{generalS}) as follows.
Denoting by $U^*_1$ and $U^*_2$ the solutions of (\ref{generall})
for $l=l_1$ and $l=l_2$, respectively, we have
\beqa \label{diffss}
S_A(l_1)-S_A(l_2) \sim  \int^\infty
dU\sqrt{\beta(U)H(U)}\left[\left(1-{H(U^*_1)\over
H(U)}\right)^{-{1\over2}} -\left(1-{H(U^*_2)\over
H(U)}\right)^{-{1\over2}}\right]
\eeqa
where we omitted an overall multiplicative constant and focused on
the behavior of the integral in the UV region $U\to\infty$. In that
region $H(U)\to\infty$, and we can approximate the integrand in
(\ref{diffss}) by
\beqa \label{approxdiff}
S_A(l_1)-S_A(l_2) \sim
(H(U^*_1)-H(U^*_2))\int^\infty dU\sqrt{\beta(U)\over H(U)}~.
\eeqa
The integrand in (\ref{approxdiff}) behaves as $U\to\infty$ in the
same way as that in (\ref{generall}). Thus, if the latter is finite
and does not require introduction of a UV cutoff, the same is
true for the former.

To find the dependence of the entropy on $l$ we need to determine
$U^*(l)$ by solving (\ref{generall}), and then use it in
(\ref{generalS}). In the next sections we will study specific
examples of this procedure; here we would like to make some general
comments on it.

Consider first the limit $U^*\to\infty$. Physically, one would
expect $l(U^*)$ to go to zero in this limit since as $l\to 0$ the
minimal action surface should be located at larger and larger $U$. In
terms of (\ref{generall}) this means that although the prefactor
$\sqrt{H(U^*)}$ goes to infinity, the integral goes to zero faster,
such that the product of the two goes to zero as well. We will see
below that this is indeed what happens in all the examples we will
consider.

It turns out that $l$ (\ref{generall}) also goes to zero in the
opposite limit $U^*\to U_0$. The prefactor $\sqrt{H(U^*)}$ goes to
zero in this limit, and as long as the integral does not diverge
rapidly enough to overwhelm it, $l\to 0$. Since any divergence of
the integral as $U^*\to U_0$ must come from the region $U\simeq
U^*\simeq U_0$, it is enough to estimate the contribution to it from
this region. In terms of the coordinate $r$ defined above, one has
\beqa \label{etsimatel} l(r_*) \sim r_*^p\int_{r_*} \frac{dr}
{\sqrt{ r^{2p} - r_*^{2p}}}~.
\eeqa
For $p>1$ one finds that for small $r_*$, $l(r_*)\sim r_*$; for
$p=1$, $l(r_*)\sim r_*\ln r_*$. In both cases, $l\to 0$ in the limit
$r_*\to 0$ (or, equivalently, $U^*\to U_0$).

We see that for small $l$ the equation of motion (\ref{firstord})
has two independent solutions, one with large $U^*$ and the other
with $U^*\simeq U_0$. The former is a local minimum of the action
(\ref{generalS}) while the latter is a saddle point. We can
interpolate between them with a sequence of curves which differ in
the minimal value of $U$, such that the solution with large $U^*$ is
a local minimum along this sequence, while the one with $U^*\simeq
U_0$ is a local maximum.

This implies that there must be another local minimum of the
effective action, with $U^*$ smaller than that of the saddle point.
This solution cannot correspond  to a smooth $U(x)$, since then it
would be captured by the above analysis. Therefore, it must
correspond to a disconnected solution, which formally has $U^*=U_0$,
but is better described as two disconnected surfaces that are
extended in all spatial directions except for $x$, and are located
at $x=\pm {l\over2}$.

The entropy corresponding to this solution is given by (see
(\ref{generalS}))
\beqa \label{disconnectedS} \frac{S_A}{V_{d-1}} =
\frac{1}{2G_N^{(10)} } \int_{U_0}^{U_\infty} dU\sqrt{\beta(U)
H(U)}~.
\eeqa
By the above analysis it must be smaller than that of the connected
solution with $U^*\simeq U_0$, but may be larger or smaller than
that of the connected local minimum with large $U^*$.

We saw before that $l(U^*)$, (\ref{generall}), goes to zero both at
large $U^*$ and as $U^*\to U_0$. If the supergravity background is
regular, one can show that between these two extremes $l$ is a
smooth function of $U^*$, that remains finite everywhere. The
simplest behavior it can have is to increase up to some point  where
$\partial l/\partial U^*=0$, and then decline back to zero as
$U^*\to\infty$. We will see that this is indeed what happens in all
the examples we study below.

Denoting the value of $l(U^*)$ at the maximum by $l_{\rm max}$, this
behavior implies that smooth solutions to the equation of motion
(\ref{firstord}) only exist for $l\le l_{\rm max}$. As $l\to l_{\rm
max}$ from below, the local minimum and saddle point discussed above
approach each other, merge and annihilate for $l>l_{\rm max}$.

At first sight, the fact that there are no solutions to
(\ref{firstord}) for $l>l_{\rm max}$ may seem puzzling, but it is
important to remember that this analysis only captures smooth
connected solutions. As discussed above, for all $l$ we have in
addition a disconnected solution for which $U'(x)$ is infinite. For
$l>l_{\rm max}$ the entanglement entropy $S_A$ is governed by this
solution and is given by (\ref{disconnectedS}). For $l<l_{\rm max}$
one needs to compare the entropies of the connected and disconnected
solutions and find the smaller one. This difference can be written
as
\beqa \label{diffS}
\frac{2G_N^{(10)}}{V_{d-1}} \left(S_A^{(\rm
conn)}-S_A^{(\rm disconn)}\right)= \int_{U^*}^\infty dU\sqrt{\beta
H}\left(\frac{1}{\sqrt{1-\frac{H(U^*)}{H(U)}}}-1\right)-
\int_{U_0}^{U^*} dU\sqrt{\beta H}~.
\eeqa
It is physically clear and easy to see from (\ref{diffS}) that for
small $l$ the connected solution with large $U^*$ has the lower
entropy. As $l$ increases, there are in general two possibilities.
The connected solution can remain the lower action one until
$l=l_{\rm max}$, or there could be a critical value $l_{\rm
crit}<l_{\rm max}$ above which the right hand side of (\ref{diffS})
is positive, so that the disconnected solution becomes the dominant
one. In the first case there would be a phase transition at
$l=l_{\rm max}$; in the second, the transition would occur at
$l_{\rm crit}$, and in the range $l_{\rm crit}<l<l_{\rm max}$, the
connected solution would be a metastable local minimum.  In all the
examples we study below it is the second possibility, $l_{\rm
crit}<l_{\rm max}$, that is realized: as we increase $l$, the
transition occurs before the connected solution ceases to exist.
This is similar to the first-order finite temperature deconfinement
transitions found in gravitational duals of confining gauge
theories \cite{Witten,Buchel,Aharony,Mahato}.

\section{D4-branes on a circle}

The low energy dynamics of $N_c$ $D4$-branes in type IIA string
theory is governed by $4+1$ dimensional supersymmetric Yang-Mills
theory with gauge group $U(N_c)$ and `t Hooft coupling
$\lambda=g_sN_cl_s$. In order to reduce to $3+1$ dimensions  and
break supersymmetry, we compactify one of the directions along the
branes, $x^4$, on a circle of radius $R_4$, $x^4\sim x^4+2\pi R_4$, with twisted boundary
conditions for the fermions.

The low energy dynamics of this system, which was studied in
\cite{Witten}, depends on the dimensionless parameter
$\lambda_4=\lambda/R_4$, and can be investigated using different
tools in different regions of parameter space. For $\lambda_4\ll 1$,
it corresponds to pure Yang-Mills theory with gauge group $U(N_c)$
and 't Hooft coupling $\lambda_4$ (at the scale $R_4$). In the
opposite limit, $\lambda_4\gg 1$, one can use a gravitational
description in terms of the near-horizon geometry of the
branes\footnote{Here and below we set $\alpha'=1$.}
\beqa\label{geomdfour} ds^2 &=& \left(\frac{U}{R}\right)^{3/2}
\left[ \left(\frac{R}{U}\right)^{3} \frac{dU^2}{f(U)} + dx^\mu
dx_\mu \right] + R^{3/2} U^{1/2} d\Omega_4^2 +
\left(\frac{U}{R}\right)^{3/2} f(U) (dx^4)^2~, \\
e^{-2 \phi} &=&   \left(\frac{R}{U}\right)^{3/2}~,
\eeqa
where $R$ is related to the five dimensional 't Hooft coupling via
the relation $R^3=\pi\lambda$, and
\beqa
 f(U) = 1 -\left( \frac{U_{0}}{U}\right)^3~, \qquad U_0 = \frac{4\pi}{9}\frac{\lambda}{R_4^2}~.
\eeqa
As $U\to U_0$, the radius of the $x^4$-circle goes to zero; $(U,
x^4)$ form together a cigar geometry of the type described in the
previous sections.

Comparing (\ref{geomdfour}) to (\ref{gengeom}) we identify $\alpha,
\beta, V_{\rm int}$ as,
\beqa\label{formback} \alpha &=& \left(\frac{U}{R}\right)^{3/2}~,
\quad
\beta = \left(\frac{R}{U}\right)^{3} \frac{1}{f(U)}~, \\
V_{\rm int} &= & \frac{8 \pi^2}{3} (R^{3} U) \times 2\pi R_4
\left(\frac{U}{R}\right)^{3/4} \sqrt{f(U)} =  \frac{32 \pi^3
R^{\frac{15}{4}} }{9 U_{0}^{\frac{1}{2}}}  U^{\frac{7}{4}}
\sqrt{f(U)}~.
\eeqa
The combination (\ref{denote}) is given in this case by
\beqa H(U) = R^6\left(\frac{32 \pi^3 }{9}\right)^2  \frac{U^2(U^3 -
U_{0}^3)} {U_{0}}~.
\eeqa
Note that $H'(U)>0$ for all $U\ge U_0$, as mentioned in the previous
section.

The explicit form of the background can be used to verify the
assertions of section 2 about the behavior of $l(U^*)$. In
particular, it is easy to check that the integral (\ref{generall})
converges. For $U^*\gg U_0$ it is given by
\beqa l(U^*)=2 R^{3/2} \times 2\sqrt{\pi}
\frac{\Gamma\left(\frac{3}{5}\right)}
{\Gamma\left(\frac{1}{10}\right)} \frac{1}{\sqrt{U^*}}~. \eeqa
We see that $l$ indeed goes to zero in the limit $U^*\to\infty$, as
expected. Similarly, one can check that it goes to zero in the
opposite limit $U^*\to U_0$. The full curve $l(U^*)$ can be computed
numerically and is plotted in figure 1. It has the qualitative
structure anticipated in section 2. The maximum of the curve occurs
at $U^*\simeq 1.2 U_0$, with
\beqa\label{lmaxdfour}
l_{\rm max}\simeq 1.418 R_4 ~.
\eeqa
At larger values of $l$, there is no smooth solution to the
equations of motion (\ref{firstord}).

\begin{figure}[ht]
\centering \label{PlotD4l}
\includegraphics{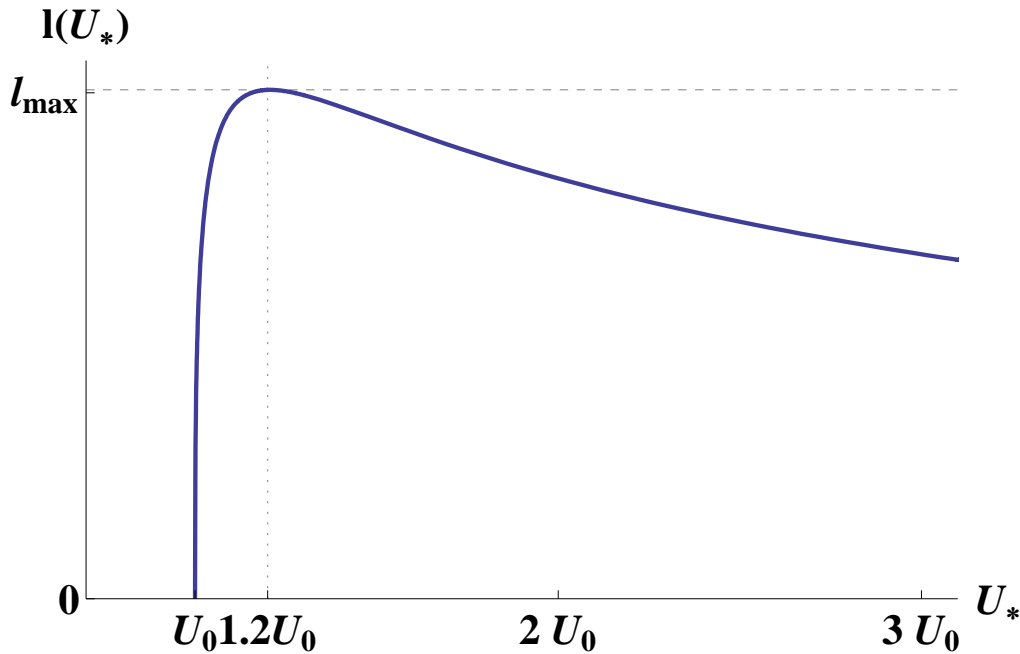}
\caption{$l(U^*)$ for $D4$-branes on a circle.}
\end{figure}

Turning to the entanglement entropy $S_A$, following the discussion
of section 2 we need to calculate the entropies of the connected
solution (\ref{generalS}) and the disconnected one
(\ref{disconnectedS}), and compare them. The calculations of the
individual entropies must be done with the UV cutoff $U_\infty$ in
place, but the difference of entropies is insensitive to it (see
(\ref{diffss}), (\ref{diffS})).

For the disconnected solution, the entropy can be calculated in
closed form:
\beqa\label{discdfour} S_A^{(\rm disconn)}  =  \frac{8\pi^3}{9}
\frac{V_2R^{9/2}}{U_{0}^{1/2}G_N^{(10)}}   \left( U_\infty^2 -
U_{0}^2 \right)~.
\eeqa
For the connected one it is given by (\ref{generalS}), which in
general has to be computed numerically. For small $l$ one can again
perform the integral using the fact that in this case $U^*\gg U_0$.
One finds
\beqa\label{conndfour} S_A^{(\rm conn)}(l) = \frac{8\pi^3}{9}
\frac{V_2R^{9/2}}{U_{0}^{1/2}G_N^{(10)}} \left(U_\infty^2  -
256 \left[\frac{\sqrt{\pi }
\Gamma\left(\frac{3}{5}\right)}{\Gamma\left(\frac{1}{10}\right)}
\right]^5
\frac{R^6}{l^4} \right)\ . \eeqa
%
Comparing to (\ref{discdfour}) we see that for small $l$
the connected solution has lower entropy, in agreement with the
general discussion of section 2. The fact that the entropy
(\ref{conndfour}) scales like $1/l^4$ at small $l$ is indicative of
$5+1$ dimensional scale invariant dynamics. This is what one
expects, since at short distances the dynamics on the wrapped $D4$-branes
is described by the $(2,0)$ superconformal field theory in $5+1$
dimensions. Indeed, we find
\beqa
 S_A^{(\rm conn)}(l)- S_A^{(\rm disconn)}
=  -V_2 (2\pi R_4)(2\pi R_{10}) \frac{32 \sqrt{\pi}}{3}
\left[\frac{\Gamma(\frac{3}{5})}{\Gamma(\frac{1}{10})}\right]^5
{N_c^3\over l^4} + \ldots \eeqa which is precisely the entanglement
entropy of $N_c$ coincident $M5$-branes compactified  on a circle of
radius $R_4$ and the M-theory circle of radius $R_{10}= g_s$ found
in \cite{Tak}.

A naive use of (\ref{discdfour}), (\ref{conndfour}) suggests that
the disconnected solution becomes the lower entropy one at $l\sim
R^{3\over2}/U_0\sim R_4$, not far from $l_{\rm max}$
(\ref{lmaxdfour}). Of course, the small $l$ approximation leading to
(\ref{conndfour}) is not valid there, and in order to determine the
precise position of the transition we need to evaluate
(\ref{generalS}). The result of that evaluation is shown in figure
2, where we also exhibit the entropy of the disconnected solution
and, for completeness, that of the saddle point discussed in section
2 as well.

\begin{figure}[ht]
\centering \label{PlotD4Sl}
\includegraphics{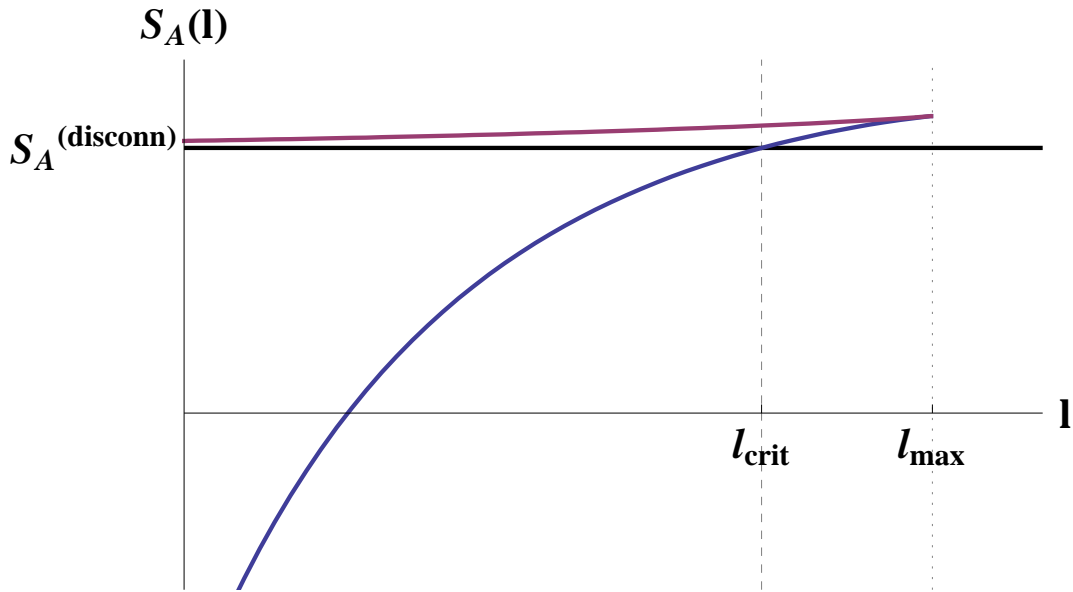}
\caption{Entropies of the connected (blue and red) and disconnected
(black) solutions for the wrapped $D4$-brane geometry.}
\end{figure}

We see that, as expected, the saddle point entropy is larger than
that of the connected and disconnected local minima for all $l$. It
approaches that of the connected one as $l\to l_{\rm max}$, and the
disconnected one as $l\to 0$. The entropies of the connected and
disconnected solutions cross at $l=l_{\rm crit}<l_{\rm max}$ given
by
\beqa\label{lcritdfour} l_{\rm crit}\simeq 1.288 R_4~.
\eeqa
As explained in section 2, the entropy is governed by the connected
solution and exhibits non-trivial dependence on $l$ for $l<l_{\rm
crit}$, while for $l>l_{\rm crit}$ it is governed by the
disconnected one and is $l$ independent (to leading order in
$1/N_c$).

\section{D3-branes on a circle}

In this section we study the system of $N_c$ $D3$-branes wrapped
around a circle of radius $R_3$ with twisted boundary conditions for
the fermions. The discussion is largely parallel to that of the
previous section, and some of the results already appear in 
\cite{Nishioka}, so we will be brief.

Before the compactification, the low energy theory on the
$D3$-branes is $N=4$ SYM with 't Hooft coupling $\lambda=g_sN_c$.
For finite $R_3$ one finds at long distances a $2+1$ dimensional
confining theory. For $\lambda\ll1$ that theory is $2+1$ dimensional
YM with 't Hooft coupling $\lambda_3=\lambda/R_3$ \cite{Witten}. For $\lambda\gg
1$ one can instead use a gravitational description in terms of the
near-horizon geometry of the $N_c$ D3-branes,
\beqa\label{dthreegeom}
 ds^2_{10} &=& \left(\frac{U}{L}\right)^2 \left[
\left(\frac{L}{U}\right)^4 \frac{dU^2}{h(U)}+dx^\mu dx_\mu \right]
+ L^2 d\Omega_5^2+ \left(\frac{U}{L}\right)^2h(U) (dx^3)^2~,  \\
h(U) &=& 1 - \left(\frac{U_0}{U}\right)^4~,
\eeqa
where
\beqa L^4=4\pi\lambda~, \qquad U_0^2={\pi\lambda\over R_3^2} ~,
\eeqa
and the dilaton is constant, $\phi(U)=0$. Comparing
(\ref{dthreegeom}) to (\ref{gengeom}) we find
\beqa \alpha = \left(\frac{U}{L}\right)^2~, \quad \beta =
\left(\frac{L}{U}\right)^4\frac{1}{h(U)}~, \quad V_{\rm int}=
 2\pi^4 R_3 L^4 U\sqrt{h(U)}~. \eeqa
The combination (\ref{denote}) is given by
\beqa H(U)= (2\pi^4 R_3)^2 L^4 U^6 h(U)\ .
\eeqa
It is again monotonically increasing with $U$, as expected.

All the calculations of the previous section can be done in this
case as well. The integral (\ref{generall}) is again convergent. For
small $l$ (and large $U^*$) one finds
\beqa
l(U^*)= 2\sqrt{\pi}  \frac{\Gamma(\frac{2}{3})}{
\Gamma(\frac{1}{6})}\frac{L^2}{U^*}~.
\eeqa
The extension to all $U^*$ is plotted in figure 3. The qualitative
shape of $l(U^*)$ is similar to the $D4$-brane case shown in figure
1. The maximum occurs at $U^* \simeq 1.113 U_0$, and
\beqa l_{\rm max}\simeq 1.383 R_3~.\eeqa
\begin{figure}[ht]
\centering \label{PlotD3l}
\includegraphics{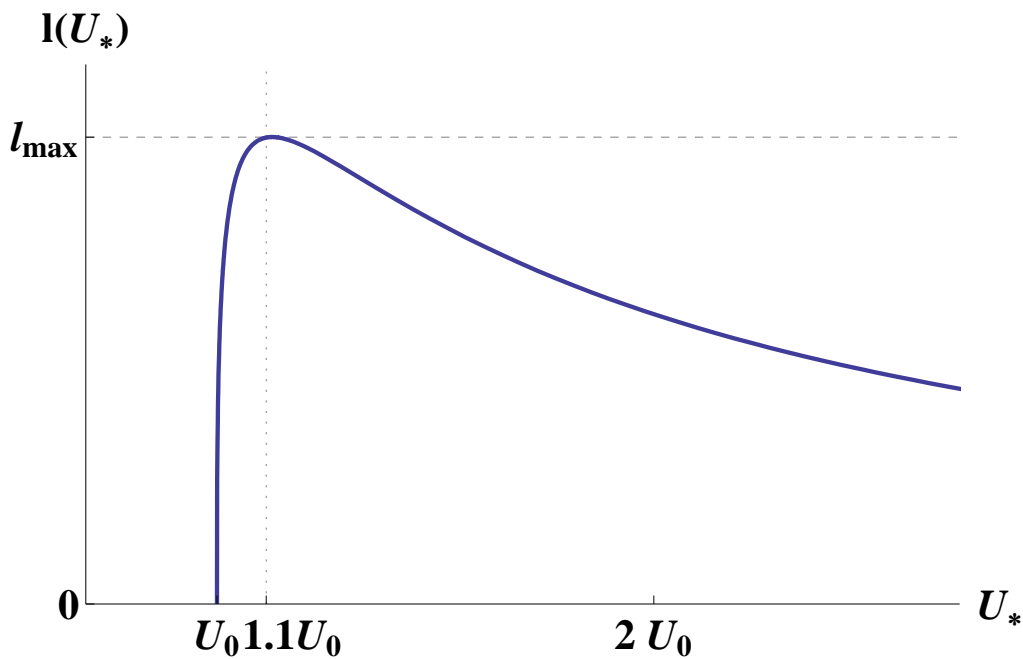}
\caption{$l(U_*)$ for $D3$-branes on a circle.}
\end{figure}
The entropy of the disconnected solution
is given by
\beqa S_A^{(\rm disconn)} = \frac{\pi^4 R_3 L^4 V_1 }{2 G_N^{(10)} } \left(
U_\infty^2 - U_0^2 \right) \ ,\eeqa
where $V_1$ is the length of the strip.
The entropy of the connected solution is exhibited in figure 4. For
small $l$ one has
\beqa S_A^{(\rm conn)}(l)= \frac{\pi^4 R_3 L^4 V_1 }{2 G_N^{(10)} } \left(
U_\infty^2 -
 4\left[\frac{\sqrt{\pi}\Gamma(\frac{2}{3})}{\Gamma(\frac{1}{6})}\right]^3
\frac{L^4}{l^2}\right)\ .\eeqa
%
Therefore, for small $l$ we find
\beqa \label{enfourent}
 S_A^{(\rm conn)}(l)- S_A^{(\rm disconn)}
=  -2 \sqrt{\pi}
\left[\frac{\Gamma(\frac{2}{3})}{\Gamma(\frac{1}{6})}\right]^3 \;
V_1(2\pi R_3) { N_c^2\over l^2} + \ldots
\eeqa
which is the entanglement entropy
of the $3+1$ dimensional ${\cal N}=4$ SYM theory compactified on a circle of radius
$R_3$ \cite{Tak}.

\begin{figure}[ht]
\centering \label{PlotD3Sl}
\includegraphics{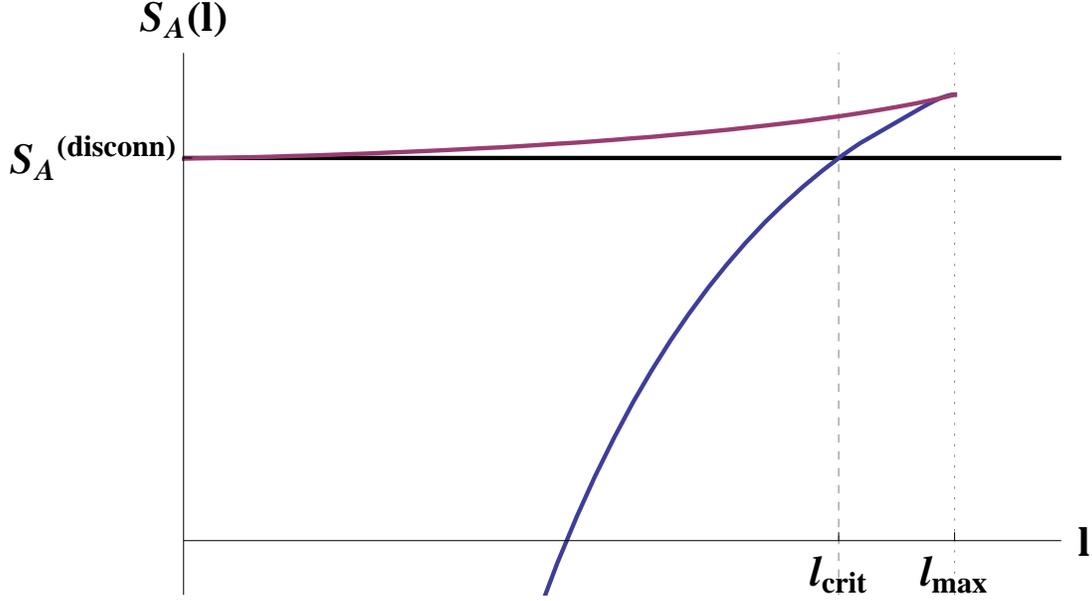}
\caption{Entropies of the connected (blue and red) and disconnected
(black) solutions for the wrapped $D3$-brane geometry.}
\end{figure}

As is clear from figure 4, the transition between the connected and
disconnected solutions happens again at a value of $l$ smaller than
$l_{\rm max}$. The numerical evaluation gives
\beqa l_{\rm crit}\simeq 1.2376 R_3~.\eeqa

\section{Cascading Confining Gauge Theory}

The background dual to the cascading $SU(M(k+1))\times SU(Mk)$
supersymmetric gauge theory is the deformed conifold $\sum_{i=1}^4
z_i^2=\epsilon^2$ warped by $M$ units of RR 3-form flux. The
relevant metric is \cite{KS},
\beq \label{specans}
ds^2_{10} = h^{-1/2}(\tau)   dx^\mu dx_\mu
 +  h^{1/2}(\tau) ds_6^2 \ ,
\eeq
where $ds_6^2$ is the metric of the deformed conifold
\beqa \label{metricd} ds_6^2 = {1\over 2}\epsilon^{4/3} K(\tau)
\Bigg[ {1\over 3 K^3(\tau)} (d\tau^2 + (g^5)^2)
 +
\cosh^2 \left({\tau\over 2}\right) [(g^3)^2 + (g^4)^2]\nonumber \\
+ \sinh^2 \left({\tau\over 2}\right)  [(g^1)^2 + (g^2)^2] \Bigg]
\ .
\eeqa
%
 Here
\beq
K(\tau)= { (\sinh (2\tau) - 2\tau)^{1/3}\over 2^{1/3} \sinh \tau}
\ ,
\eeq
and the warp factor is given by
\beq \label{intsol}
h(\tau) =
(g_s M\alpha')^2 2^{2/3} \epsilon^{-8/3}
\int_\tau^\infty d x {x\coth x-1\over \sinh^2 x} (\sinh (2x) - 2x)^{1/3}
\ .
\eeq
The dilaton is constant and we set it to zero.
For the details of the angular forms $g_i$, see \cite{KS,HKO}.

The cascading gauge theory has a continuous parameter, $g_s M$. The
theory approaches the pure $SU(M)$ SYM theory in the limit $g_s M\to
0$, while the dual supergravity description is reliable in the
opposite limit, $g_s M\to \infty$. In this limit the geometry
describes a gauge theory with two widely separated scales: the scale
of glueball masses,
\beqa \label{mglue}
m_{\rm glueball} = \frac{\epsilon^{2/3}}{ g_s M
\alpha'} \ ,
\eeqa
and the scale of the string tension at the IR end of space (the tip
of the cigar), $\sqrt{T_s}\sim \sqrt{g_s M} m_{\rm glueball}$.

The metric (\ref{specans}), (\ref{metricd}) is of the form
(\ref{gengeom}) with
\beq \label{combo} \alpha \equiv h^{-1/2} , \quad \beta \equiv
\frac{h(\tau) \epsilon^{4/3}}{6 K^2(\tau)}~.
\eeq
Using $\int g_1\wedge g_2\wedge g_3\wedge g_4\wedge g_5 = 64 \pi^3$, we get
\beq V_{\rm int} = {4 \pi^3 \over \sqrt{6}} \; h^{5/4} \epsilon^{10/3} K \sinh^2(\tau)
\ .\eeq
Thus, all the general formulae of section 2 apply, with $U$ replaced by the
standard deformed conifold radial variable $\tau$.

\begin{figure}[ht]
\centering \label{PlotKSl}
\includegraphics{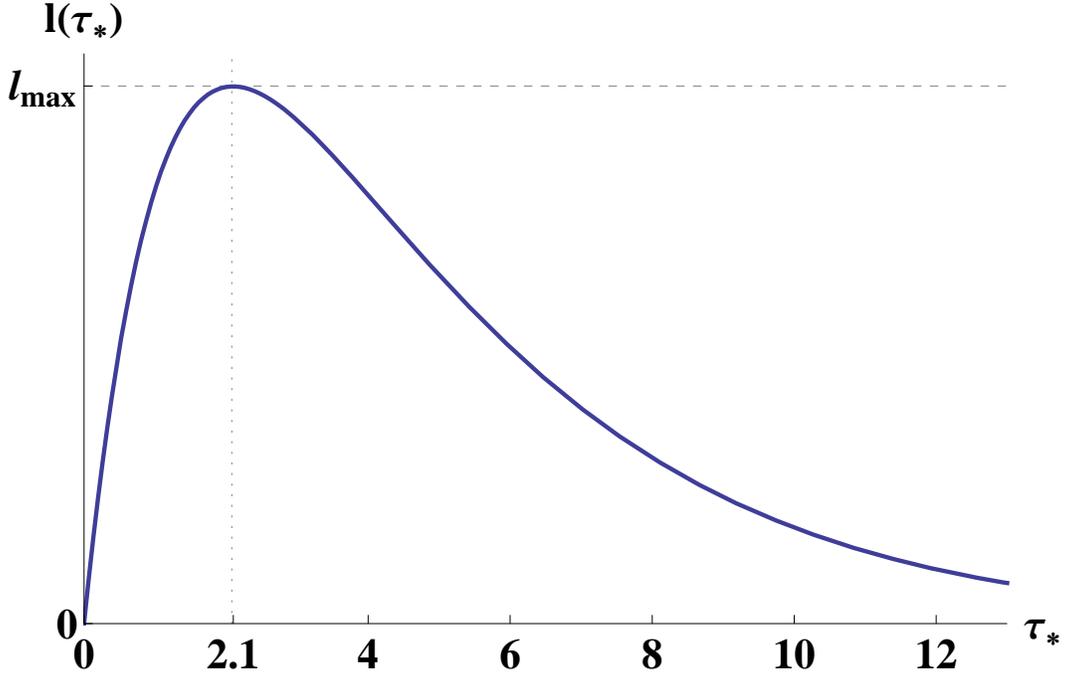}
\caption{$l(\tau_*)$ for the KS geometry.}
\end{figure}

We find \beqa H (\tau)= e^{-4 \phi} V_{\rm int}^2 \alpha^{3} =
{8\pi^6\over 3} \epsilon^{20/3} h(\tau) K^2(\tau) \sinh^4(\tau) \ .
\eeqa $H$ can be seen to be monotonically increasing with $\tau$ as
noted in section \ref{Generalsection} from general considerations.
The general equation (\ref{generall}) with these identifications
gives $l(\tau_*)$ for the KS background. As in the previous
sections, the integral is convergent. For large $\tau^*$, we can
approximate $l(\tau)$ using the asymptotic forms valid at large
$\tau$,
\beqa
h(\tau) &\to& 2^{1/3} 3 \; (g_s M \alpha')^2 \epsilon^{-8/3} \;\;
\left (\tau-\frac{1}{ 4}\right) e^{- 4 \tau/3}~, \quad \quad
K \to 2^{1/3} e^{-\tau/3}~, \\
H(\tau) &\to& \pi^6 \epsilon^{4} ( g_s M \alpha')^2 \;\; \left (\tau-\frac{1}{ 4}\right)
e^{2\tau}~, \quad \quad \sqrt{\beta} \to 2^{-2/3} \epsilon^{-2/3}
(g_s M \alpha') \sqrt{\tau-\frac{1}{4} } e^{-\tau/3}~.
\eeqa
This leads to the simplified expression,
\beqa l(\tau^*) = 2^{1/3} \epsilon^{-2/3} g_s M \alpha'
\int_{\tau^*}^\infty \frac{ \sqrt{\tau} e^{-\tau/3}   d\tau}{ \sqrt{
\frac{\tau e^{2 \tau}}{\tau^* e^{2 \tau^*} } -1 } }~.
\eeqa
The main contribution is from the region $\tau \sim \tau^*$;
shifting $\tau \to \tau_* + y$ and keeping the lowest order term in
$y$ we conclude that for large $\tau^*$,
\beqa l(\tau^*) = \frac{2^{1/3} 3 \sqrt{\pi}
\Gamma(2/3)}{\Gamma(1/6)} \epsilon^{-2/3} g_s M \alpha'
\sqrt{\tau^*} e^{- \tau^*/3}
\eeqa
As earlier, $l$ goes to zero as $\tau^* \to \infty$. One can also
verify, as outlined in section \ref{Generalsection}, that as $\tau^*
\to 0$, $l$ goes to zero again. The full curve, computed
numerically, is presented in figure \ref{PlotKSl}. We see that it
shows the same qualitative behavior as the other cases (figures
1,3). The maximum occurs at $\tau^* \approx 2.1$ with
\beqa
l_{\rm max} \approx 1.00\;\; m_{\rm glueball}^{-1} \ .
\eeqa

We now turn to the entanglement entropy $S_A$. As earlier, we have
to calculate and compare the entropies of the connected
(\ref{generalS}) and disconnected (\ref{disconnectedS}) surfaces. As
discussed in section 2, each of these entropies must be computed
with a UV cut-off in place, but the difference of the entropies is
UV finite. The entropy of the disconnected solution is found to be
%
\beqa
S_A^{(\rm disconn)} & = &
V_2 \frac{M^2 \epsilon^{4/3}}{ 2^{2/3}  16\pi^3 \alpha'^2}
\; \left( \frac{3}{2}
\tau_\infty e^{2 \tau_\infty/3} - \frac{21}{8} e^{2
\tau_\infty/3}\;\; + \;\; 2.194 \right)
\eeqa
where the finite additive constant was computed numerically.
For the connected solution, we first consider an analytic
approximation valid for small $l$:
\beqa S_A(\tau^*) = V_2 \frac{M^2 \epsilon^{4/3}}{ 2^{2/3}  16\pi^3 \alpha'^2}
\;\;   \; \int_{\tau^*}^{\tau_\infty} \frac{
(\tau-1/4)^{3/2} e^{5\tau/3}   d\tau}{ \sqrt{ (\tau-1/4) e^{2 \tau} -
(\tau^* -1/4)
e^{2 \tau^*}  } } \ .\eeqa
Approximating this integral as we did for $l(\tau^*)$, we find
\beqa S_A(\tau^*) = V_2 \frac{M^2 \epsilon^{4/3}}{ 2^{2/3}  16\pi^3 \alpha'^2}
\;\; M^2 \epsilon^{4/3} \;  \left( \frac{3}{2} \tau_\infty e^{2
\tau_\infty/3} - \frac{21}{8} e^{2 \tau_\infty/3} - \frac{3
\sqrt{\pi} \Gamma(2/3) }{2 \Gamma(1/6 )} \tau^* e^{2 \tau^*/3}
\right) \ .
\eeqa
Thus, for $l\ll 1/m_{\rm glueball}$,
\beqa S_A^{(\rm conn)}- S_A^{(\rm disconn)} = - V_2
\frac{243  \Gamma\left(\frac{2}{3}\right)^3} {32 \pi^{3/2}
\Gamma\left(\frac{1}{6}\right)^3}  \;\;   \; \frac{g_s^2
M^4}{l^2}\log^2 (m_{\rm glueball} l)+ \ldots\ . \eeqa
In the cascading theory the effective number
of colors is a logarithmic function of the distance scale \cite{KT,KS,HKO}:
\beq N_{\rm eff}(l) = {3\over 2\pi} g_s M^2 \log (m_{\rm glueball} l)+ \ldots ~.
\eeq
We see that the finite piece of the entropy is
\beq \label{cascadeent}
- V_2 \frac{27 \sqrt{\pi}   \Gamma\left(\frac{2}{3}\right)^3 N_{\rm eff}^2(l) } {8
\Gamma\left(\frac{1}{6}\right)^3 l^2}  + \ldots\ .
\eeq
For a $3+1$
dimensional conformal gauge theory, the finite piece of the entanglement
entropy is indeed of the form $N_c^2 (V_2/l^2)$.
Following \cite{Tak}, we may use a minimal surface in $AdS_5\times T^{11}$
to find the entanglement entropy in the dual
$SU(N)\times SU(N)$ SCFT \cite{KW}:\footnote{The extra factor of $27/16$ compared to the
result (\ref{enfourent}) for
$AdS_5\times S^5$ comes from the fact that ${\rm vol} (T^{11})= \frac{16}{27} {\rm vol} (S^5)$.}
 \beq
- V_2 \frac{27 \sqrt{\pi}   \Gamma\left(\frac{2}{3}\right)^3 N^2} {8
\Gamma\left(\frac{1}{6}\right)^3 l^2}
\ .\eeq
Hence, the result (\ref{cascadeent}) we find for the cascading gauge theory is a
reasonably modified form of the
conformal behavior. The same distance-dependent effective number of
colors was found in evaluation of correlation functions in the
cascading theory \cite{Krasnitz,OBY}.

\begin{figure}[ht]
\centering \label{PlotKSSl}
\includegraphics{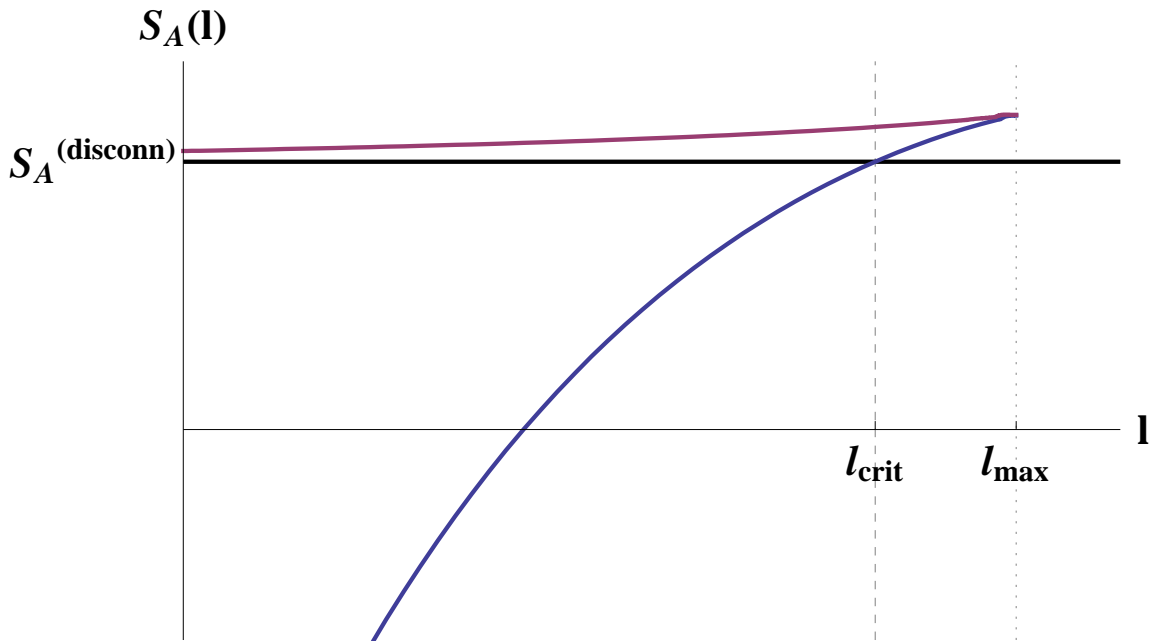}
\caption{Entropies of the connected (blue and red) and disconnected
(black) solutions for the KS geometry.}
\end{figure}

Going beyond the small $l$ limit, we present the result of the
numerical evaluation of $S_A$ in figure \ref{PlotKSSl}, which
compares the connected, disconnected and saddle point entropies. As
expected, the saddle point entropy is always the largest and
approaches the disconnected solution for small $l$ and the connected
solution as $l \to l_{\rm max}$. The connected solution has the
lowest entropy for small $l$ and is the dominant contribution in
this regime. The point at which the connected and disconnected
solutions cross is $l_{\rm crit} < l_{\rm max}$, which is found to
be
\beqa
l_{\rm crit} \approx   0.95 \;\; m_{\rm glueball}^{-1}~.
\eeqa
For $l > l_{\rm crit}$, the $O(N_c^2)$ entropy is $l$-independent as
explained in section \ref{Generalsection}.

\section{Comparison to field theory}
\label{GlueballSection}

It is natural to ask whether the transition at finite $l$ that we
found in confining gravitational backgrounds also occurs in large
$N_c$ asymptotically free gauge theories, such as pure YM or ${\cal
N}=1$ SYM with gauge group $SU(N_c)$. The location of such a
transition would have to be around the QCD scale, $l_{\rm
crit}\Lambda_{QCD}\sim 1$. At such scales the theory is strongly
coupled and it is difficult to evaluate the entanglement entropy
$S_A$ directly.

To proceed one can use the fact that, at large $N_c$, confining
gauge theories are expected to reduce to free field theories of
glueballs, whose density of states grows like
\beqa\label{densglue} \rho(m)\simeq m^\alpha e^{\beta_Hm} \eeqa
at large mass $m$. The inverse Hagedorn temperature $\beta_H$ is of
order $1/\Lambda_{QCD}$, and $\alpha$ is a constant. Both are
difficult to calculate from first principles. Most of the states
that contribute to (\ref{densglue}) are unstable resonances whose
width goes to zero as $N_c\to\infty$. More generally, all
interactions between the glueballs go to zero in this limit. At
finite $N_c$ the spectrum (\ref{densglue}) is effectively cut off at
some large mass scale.

We can use the above picture to calculate the entanglement entropy
at large $N_c$, by summing the contributions of the glueballs. To
avoid UV divergences, we will consider the quantity
\beqa\label{cdef}
C = l \frac{dS_A(l)}{dl}  \eeqa
which, as mentioned in the previous sections, does not depend on the
UV cutoff. Consider, for example, a free scalar field of mass $m$.
It is clear that the non-trivial dependence of (\ref{cdef}) on $l$
is via the combination $ml$. We will be interested in the region
$ml\gg 1$, where $C(m l)$ can be calculated as follows. In $1+1$
dimensions, the large $l$ form of $C(m l)$ has been obtained in
\cite{Casini:2005zv}; it is given by
\beqa\label{scalarc} C_1(ml)={ml\over4}K_1(2ml)\simeq {\sqrt{\pi
ml}\over8}e^{-2ml} \eeqa
A four-dimensional free scalar field can be thought of as an
infinite collection of two-dimensional ones, labeled by the momentum
in the transverse $\IR^2$, $\vec k$, with mass $m(\vec
k)=\sqrt{m^2+\vec k^2}$. Summing over these momentum modes we find
the $3+1$ dimensional version of (\ref{scalarc}),
\beqa\label{fourdsc} C_3(m l)={V_2\over(2\pi)^2}\int d^2\vec
kC_1(m(\vec k)l)\simeq \frac{V_2}{32 \pi} {\sqrt{\pi} m^2\over
\sqrt{ml}} e^{-2ml} ~. \eeqa
We see that the contribution of a single  scalar field to the
entanglement entropy is exponentially suppressed at large
mass.\footnote{The same is true for fermions and other higher spin
fields.} This is similar to the exponential suppression of its
contribution to the canonical partition sum at finite temperature,
with the role of the inverse temperature $\beta$ played here by
$2l$.

For a theory with a Hagedorn spectrum (\ref{densglue}) of bound
states, the total entropy is obtained by summing over all states,
\beqa\label{ctotal}
C_{\rm total}=\int dm \rho(m) C_3(m)
\sim\int^\infty dm m^\beta e^{(\beta_H-2l)m}
\eeqa
The integral converges for $l>\beta_H/2$ and diverges otherwise.
This is the analog of the usual Hagedorn divergence of the canonical
partition sum at the Hagedorn temperature. There, the physical
picture is that for temperatures below some critical temperature,
that is believed to be somewhat below the Hagedorn one \cite{Atick,Barbon},
the system
is in the confining phase and the thermal free energy scales like
$N_c^0$. Above that temperature, the system is in a deconfined phase
and the free energy scales like $N_c^2$.

Similarly, for the entanglement entropy in gauge theory we expect
that for $l$ above some $l_{\rm crit}$ that is somewhat larger than
$\beta_H/2$ the entanglement entropy is of order $N_c^0$ and is
given by the convergent integral (\ref{ctotal}), while for $l<l_{\rm
crit}$ the entropy is of order $N_c^2$, in agreemeent with the
divergence of (\ref{ctotal}).

The resulting picture is qualitatively similar to what we got from
the gravity analysis in sections 2 -- 5. Of course, as usual, the
details are expected to differ because in the gravity regime the
theory contains two widely separated scales. One is the scale of the
lightest glueball masses, which goes like $1/R_4$ in the $D4$-brane
analysis of section 3, like $1/R_3$ in that of section 4, and like
$m_{\rm glueball}$ (\ref{mglue}) in the KS geometry. The other is
the scale of massive string excitations living near the tip of the
cigar, $\sqrt{T_s}$, which is parametrically higher than the
glueball scale. Since the exponential density of states comes from
these string modes, we expect $\beta_H$ to be of order $T_s^{-1/2}$.

The transition point $l_{\rm crit}$ in the gravity regime is instead
determined by the inverse of the lightest glueball mass, and is
parametrically larger than the Hagedorn scale $T_s^{-1/2}$. Thus, as
we decrease $l$, the transition at $l=l_{\rm crit}$ to entangelement
entropy of order $N_c^2$ happens long before $\beta_H$, $l_{\rm
crit}\gg\beta_H$. For example, in the cascading theory $l_{\rm
crit}/\beta_H \sim \sqrt{g_s M}$.

In the asymptotically free field theory regime, there is a single
scale $\Lambda_{QCD}$ and everything happens around it. One can
interpolate between the two regimes by tuning the 't Hooft coupling
(e.g. making $g_s M$ small in the KS example). Our results suggest
that no phase transition is encountered along such an interpolation
-- the two regimes are in the same universality class.

The arguments presented above apply directly to large $N_c$
theories. It would be interesting to investigate whether the phase
transition we found continues to exist at finite $N_c$, and to
characterize its order. Studying the entanglement entropy in pure
glue $SU(N_c)$ lattice gauge theory would therefore be very
interesting.

\section{Discussion}

In this paper we applied the holographic method for calculating the
entanglement entropy, introduced in \cite{Tak}, to confining
theories with gravity duals. In the simple case of entanglement
between a strip of width $l$ and its complement, we found an
interesting phase transition as a function of $l$: for $l< l_{\rm
crit}$ the entropy is dominated by the action of a connected
surface, while for $l> l_{\rm crit}$ by that of a disconnected one.
After a subtraction of an $l$-independent UV divergent contribution,
we conclude that the entropy is $O(N_c^2)$ for $l< l_{\rm crit}$ and
$O(1)$ for $l> l_{\rm crit}$. This transition is qualitatively
similar to the confinement/deconfinement transition at finite
temperature.

Studying the thermal phase transition in confining gravitational
backgrounds requires finding a SUGRA solution with an event horizon,
and comparing its action with that of another solution which is
horizon-free but has the Euclidean time periodically identified
\cite{Witten}. In general, these calculations are complicated and
require a considerable amount of numerical work (see, for example,
\cite{Buchel,Aharony,Mahato}). Studying the qualitatively similar
transition for the entanglement entropy is much simpler; instead of
finding new SUGRA solutions, one needs to find locally stable
surfaces in previously known backgrounds.

We also argued that a transition similar to the one we observed
using the methods of \cite{Tak} should occur in any confining large
$N_c$ gauge theory. This reasoning, and the several examples we have
presented, make it plausible that any consistent gravity dual of a
confining theory has to exhibit this phase transition. This is a
useful prediction for any confining gauge/gravity dual pairs that
remain to be discovered.

The existence of the transition in the cases we have discussed is
linked to $p$-cycles of the internal geometry that shrink in the IR.
One could ask if this is the most general situation that results in
the phase transition. As we showed, the monotonic function $H(U) =
e^{-4 \phi} V_{\rm int}^2  \alpha^d$ has to vanish at the IR ``end
of space,'' $U=U_0$. On the other hand, $\alpha(U_0)$ should be
non-vanishing for the string to retain its tension in the IR. This
seems to restrict us to the vanishing of $e^{-2\phi} V_{\rm int}$.
Thus, we should consider models where there are shrinking cycles
and/or $\phi$ diverges in the IR.

Curiously, one of the most widely used gravitational models of
confinement \cite{Polchinski}, $AdS_5$ with a hard IR wall at
$U=U_0$, exhibits neither of these phenomena because both $\phi$ and
$V_{\rm int}$ are assumed to be constant. Therefore, for such a
model the transition of the entanglement entropy does not seem to
occur. This is not surprising, since the notion of the disconnected
solution wrapping the entire geometry is not {\it a priori}
well-defined in this case. A related problem is that the equations
of motion are not satisfied at $U=U_0$, hence the boundary
conditions are ambiguous there.

There may exist a definition of the boundary conditions that allows
the disconnected solution and produces a phase transition of the
entanglement entropy (an encouraging sign is that the thermal
deconfining transition does take place in the hard-wall model
\cite{Herzog}). Indeed, when the hard wall model was considered in
\cite{Tak} the contribution from the part of the minimal surface
lying along the hard IR wall was not included in the calculation;
hence, it was treated as a disconnected surface.
Justifying such a prescription may be a good problem for the future.

Another popular phenomenological model is the ``soft wall'' model
where space-time has the geometry of $AdS_5$, while $\phi(U)$ blows
up in the IR \cite{Karch}:
\beqa \label{softwall}
ds_5^2 = U^2 \left (U^{-4} dU^2 + dx^\mu
dx_\mu \right )\ , \qquad \phi(U) =U^{-2}\ .
\eeqa
Here, there is no shrinking internal cycle but the blow-up of the
dilaton causes $H(U)$ to rapidly approach zero at $U=0$.\footnote{
For the soft-wall model $\alpha(U)=U^2$, hence the string loses its
tension at $U=0$. However, the model is typically treated as a five
dimensional field theory, so it is not clear if the string tension
requirement needs to be imposed.} In general, if $H(U)\sim U^p
e^{-k/U^q}$ as $U\to0$, one finds  a finite $l_{\rm max}$ (above
which the connected solution does not exist) provided $\beta(U)$ has
a pole of order $2 q + 2$ or less at $U=0$. One can show this by
similar means to those employed in section \ref{Generalsection}
where only shrinking cycles were considered. In all the models
considered in this paper so far, $q=0$ and $\beta$ had a pole of
order less than 2. On the other hand, the soft wall model
corresponds to $q=2$ while $\beta(U) = 1/U^4$ and hence still
satisfies the criterion for the existence of a finite $l_{\rm max}$.
For the soft-wall model one finds that there is indeed a transition
between the disconnected solution stretching from $U=0$ to
$U=\infty$ and the connected one that becomes unstable for $l>
l_{\rm crit}$ and stops existing at $l_{\rm max}$.

We see that the entanglement entropy may be useful as a simple test
of holographic models of confinement. More amibitiously, it would be
nice to show that, if the confining background satisfies the
supergravity equations of motion (neither the hard-wall nor the
soft-wall do), then there is a phase transition of the entanglement
entropy.

Finally, it is important to understand the underlying reasons for
the success of the geometric method of \cite{Tak}. This prescription
is designed to capture only the leading, $O(N_c^2)$, term in the
entanglement entropy. While it has a superficial similarity to probe
brane calculations, it does not seem to be consistent to think of
the bulk surface that appears in the construction as a brane. Indeed
a brane with the worldvolume action (\ref{prescription}) would have
tension proportional to $1/g_s^2$, and would back-react on the
geometry at leading order in $g_s$. In any case, branes with the
right properties do not seem to exist (see e.g. \cite{ExoticSeven}).
We need to formulate the problem in semiclassical gravity whose
solution to leading order in $G_N^{(10)}$ is the minimization
problem proposed in \cite{Tak}. Hopefully, this can pave the way to
finding the $O(N_c^0)$ corrections to the entanglement entropy and
comparing them with field theory.

\section*{Acknowledgments}
We are grateful to Marcus Benna, Oleg Lunin, Juan Maldacena, Dmitry Malyshev and Tadashi Takayanagi for
useful discussions. I.K. acknowledges the hospitality of the University of Tokyo, Komaba, and the Aspen
Center for Physics, where some of his work on this project was
carried out. A.M. and D.K. acknowledge the hospitality of TASI,
Boulder where some of this work was carried out. The work of I.K. and A.M. was
supported in part by the National Science Foundation under Grant No.
PHY-0243680.
The work of D.K. is supported in part by the Department of Energy under grant DE-FG02-90ER40560,
the National Science Foundation under grant 0529954 and the Joint Theory Institute funded by Argonne
National Laboratory and the University of Chicago.
Any opinions, findings, and
conclusions or recommendations expressed in this material are those
of the authors and do not necessarily reflect the views of these
funding agencies.

\bibliographystyle{unsrt}

\end{document}